# Ultralow-loss domain wall motion driven by magnetocrystalline anisotropy gradient in antiferromagnetic nanowire

D. L. Wen[1,*], Z. Y. Chen[1,*], W. H. Li[1], M. H. Qin[1,†], D. Y. Chen[1], Z. Fan[1], M. Zeng[1], X. B. Lu[1], X. S. Gao[1], and J. –M. Liu[1,2]

[1]*Institute for Advanced Materials, South China Academy of Advanced Optoelectronics and Guangdong Provincial Key Laboratory of Quantum Engineering and Quantum Materials, South China Normal University, Guangzhou 510006, China*

[2]*Laboratory of Solid State Microstructures, Nanjing University, Nanjing 210093, China*

[Abstract] Searching for new methods controlling antiferromagnetic (AFM) domain wall is one of the most important issues for AFM spintronic device operation. In this work, we study theoretically the domain wall motion of an AFM nanowire, driven by the axial anisotropy gradient generated by external electric field, allowing the electro control of AFM domain wall motion in the merit of ultra-low energy loss. The domain wall velocity depending on the anisotropy gradient magnitude and intrinsic material properties is simulated based on the Landau-Lifshitz-Gilbert equation and also deduced using the energy dissipation theorem. It is found that the domain wall moves at a nearly constant velocity for small gradient, and accelerates for large gradient due to the enlarged domain wall width. The domain wall mobility is independent of lattice dimension and types of domain wall, while it is enhanced by the Dzyaloshinskii-Moriya interaction. In addition, the physical mechanism for much faster AFM wall dynamics than ferromagnetic wall dynamics is qualitatively explained. This work unveils a promising strategy for controlling the AFM domain walls, benefiting to future AFM spintronic applications.

---
[*]D Wen and Z Chen contributed equally to this work;
[†]E-mail: qinmh@scnu.edu.cn

# I. Introduction

Nowadays, the interest in antiferromagnets significantly increases due to the promising application potentials for spintronics.[1,2] Comparing with ferromagnets based storage devices, antiferromagnets based devices are more stable against magnetic field perturbations and could be designed with high element densities without producing any stray field, attributing to zero net magnetization and ultralow susceptibility of an antiferromagnetic (AFM) element.[3-6] Moreover, AFM materials show fast magnetic dynamics,[7] including the spin wave modes with high frequencies resulting from their complex spin configurations, which also favors them for promising potentials in future devices.[8] One of the major challenges for these applications is how to manipulate an AFM domain wall. A number of driving schemes including spin waves,[9] spin-orbit torques,[10,11] asymmetric magnetic fields,[12] and thermal gradients[13-15] have been proposed. For example, a wall motion speed as high as ~30 km/s, driven by electric current induced Néel spin-orbit torques, available in CuMnAs[16] and $Mn_2Au$[17], was predicted. This speed is likely three orders of magnitude larger than that for ferromagnetic (FM) domain wall motion typically, although electric current driving would be highly energy-costing. Furthermore, it was revealed that the competition between entropic torque and Brownian force under temperature gradient determines the wall motion direction. Importantly, in these control schemes, the AFM wall motion remains non-tilted and high wall mobility is expected since no walker breakdown[18,19] which limits the motion of typical FM domain wall motion is available in the AFM wall motion.

While these proposed schemes are meaningful for future experiments and device design, several shortcomings can be found and they may be detrimental in some cases. For instance, the Néel spin-orbit torque is available only in some specific antiferromagnets with locally broken inversion symmetry. This torque however drives the neighboring domain walls to approach to and annihilate with each other, seriously hindering its applications. Some schemes, such as temperature gradient and spin wave driven wall motion, seem to be schemes of more theoretical sense and will encounter difficulties practically. It is also noted that some of these schemes rely essentially on the electric current driven mechanism, which is highly deficient for its high energy loss due to the Joule heating. Along this line, it is still on the way to find alternative scheme that is well-controlled, energy-saving, and high efficient for future

AFM spintronic applications.

First, electric field control rather than electric current driving would be highly preferred if one is concerned with the energy cost which is detrimental for ultra-density memories. Indeed, voltage control or say electro-control of magnetism has been an issue deserved for full commitment. Fortunately, the electric field controlled magnetic anisotropy has been experimentally revealed in magnetic heterostructures,[20-23] and the anisotropy gradient can be obtained through elaborate structure design. Under such a gradient, the AFM domain wall tends to move towards the low anisotropy side in order to reduce free energy. As a matter of fact, the anisotropy gradient has been proven to efficiently drive the skyrmions motion,[24-26] and this scheme should be also utilized to control the AFM domain wall motion with preferred merit. However, so far no report on the dynamics of AFM domain wall under such anisotropy gradient is available, and this subject urgently deserves to be investigated theoretically as the first step towards promoting the application process for AFM spintronics.

In this work, we study the anisotropy gradient driven AFM domain wall motion in a nanowire with the AFM domain structure. The dynamics and its dependence of the gradient magnitude and intrinsic physical parameters are simulated based on the Landau - Lifshitz - Gilbert (LLG) equation and also calculated based on the energy dissipation theorem. It is predicted that the domain wall does shift at a nearly constant speed under small anisotropy gradient, and can be accelerated under large anisotropy gradient because of the widened domain wall during the motion stage. It is confirmed that the observed phenomena are independent of the lattice dimension and wall type. Moreover, the Dzyaloshinskii-Moriya (DM) interaction is discussed specifically and its influence on spin dynamics is addressed, benefiting a lot to future materials design and dynamic manipulation. In addition, the physical mechanism of the AFM wall dynamics faster than FM wall dynamics is also qualitatively explained.

**II. Model and computational details**

Without losing the generality, we study a one-dimensional AFM model with the isotropic Heisenberg exchanges between the nearest neighbors and the uniaxial anisotropy term:[27]

$$H = J \sum_{<i,j>} S_i \cdot S_j - \sum_i K_i^z (S_i^z)^2, \qquad (1)$$

where $J > 0$ is the AFM coupling constant, $\boldsymbol{S}_i = \boldsymbol{\mu}_i/\mu_s$ represents the normalized magnetic moment at site $i$ with three components $S_i^x$, $S_i^y$ and $S_i^z$. The second term in Eq. (1) is the anisotropy energy with the easy axis along the $z$-direction (i.e. nanowire axis), and the uniaxial anisotropy constant $K_i^z$ at site $i$ is described by $K_i^z = K_0 + ia \cdot \Delta K$ where $K_0$ is the anisotropy constant at the low anisotropy end and $\Delta K$ describes the anisotropy gradient magnitude, $a$ is the lattice constant.

The AFM dynamics is investigated by solving the LLG equation based on the atomistic spin model:[28,29]

$$\frac{\partial S_i}{\partial t} = -\frac{\gamma}{\mu_s(1+\alpha)^2} S_i \times [H_i + \alpha(S_i \times H_i)], \qquad (2)$$

where $\gamma$ is the gyromagnetic ratio, $\alpha = 0.002$ is the Gilbert damping constant, $H_i = -\partial H/\partial S_i$ is the effective field. Unless stated elsewhere, the LLG simulations are performed on a one-dimensional nanowire lattice with $1 \times 1 \times 400$ spins with open boundary conditions using the fourth-order Runge-Kutta method with a time step $\Delta t = 1.0 \times 10^{-4}\ \mu_s/\gamma J$. The local staggered magnetization $2\boldsymbol{n} = \boldsymbol{m}_1 - \boldsymbol{m}_2$ is calculated to describe the spin dynamics where $\boldsymbol{m}_1$ and $\boldsymbol{m}_2$ are the magnetizations of the two sublattices. After sufficient relaxation of the domain structure, the anisotropy gradient is applied to drive the domain wall motion, as schematically depicted in Fig. 1.

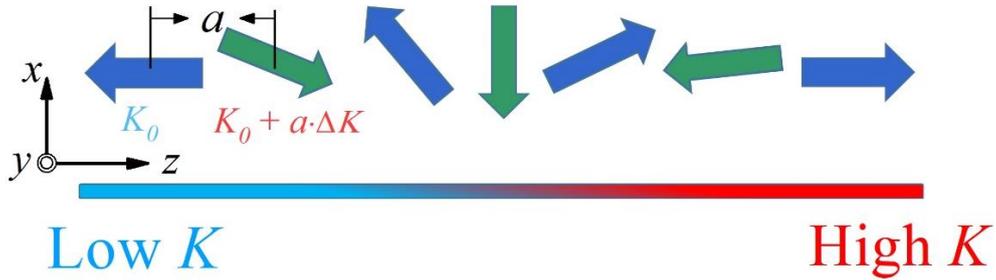

Fig.1. (color online) Illustration of a domain wall in antiferromagnetic nanowire under an anisotropy gradient.

### III. Results and discussion

*A. Domain wall motion*

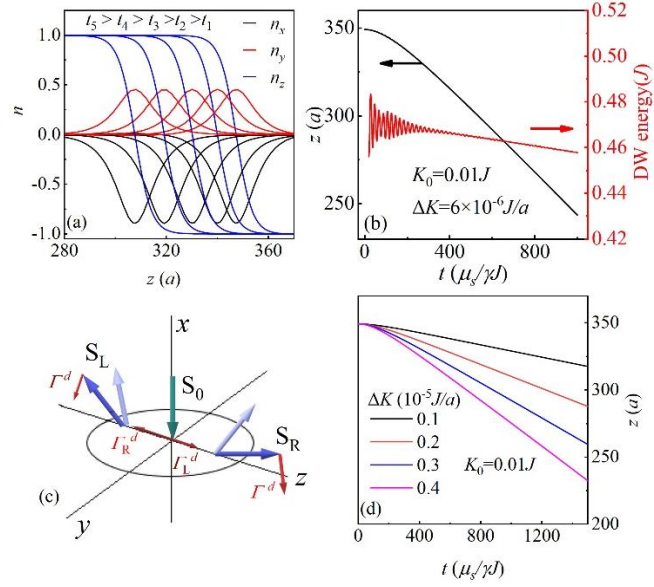

Fig.2. (color online) (a) Profile of the domain wall at different times, and (b) the domain wall position and energy as functions of time for $\Delta K = 6\times10^{-6}\ J/a$ and $K_0 = 0.01\ J$. (c) A schematic depiction of torques acting on domain wall spins under the anisotropy gradient, and (d) the position of the domain wall versus time for various $\Delta K$ for $K_0 = 0.01\ J$.

We first demonstrate the domain wall motion upon a finite anisotropy gradient $\Delta K$. It is clear to see no identifiable wall motion at $\Delta K = 0$. The motion trajectory at $K_0 = 0.01J$ and $\Delta K = 6 \times 10^6\ J/a$, at five given times, are presented in Fig. 2(a) where the domain wall profiles are plotted. The domain wall motion in roughly steady state is clearly demonstrated as long as a finite anisotropy gradient is imposed. The wall always shifts from the high anisotropy side to the low anisotropy side, with a nearly constant speed. Moreover, the wall profile is rather robust and remains in its initial configuration during the motion. The simulated wall position ($z$) and wall energy ($E_{DW}$, the internal energy difference between the systems with and without the wall) as functions of time $t$ are shown in Fig. 2(b). It is seen that the wall position $z$ decreases very slowly in the very beginning period and then the wall motion is accelerated up to a state with roughly constant speed. The wall energy $E_{DW}$ however experiences serious fluctuations in the beginning period, and then reaches a state where $E_{DW}$ approximately linearly decreases with time $t$ plus very weak fluctuations. It is thus clearly demonstrated that the domain wall motion towards the low anisotropy side is a spontaneous process. In the beginning period, the wall symmetry is broken slightly by introducing the anisotropy gradient, resulting in the distortion of the wall structure and serious fluctuation of $E_{DW}$. Moreover, the trajectory of **n** depends not only on its instantaneous position but also on the spin dynamics.[30] Specifically, **n** tends to continue its path even after reaching the equilibrium position (along its effective field), accounting for the weak fluctuations of $E_{DW}$ after the wall structure was readjusted.

As a matter of fact, the wall motion can be qualitatively understood from the competition among various torques acting on the wall spins, as depicted in Fig. 2(c). In the absence of anisotropy gradient, the domain wall configuration is symmetric with respect to its central plane. This symmetry is broken if one considers a nonzero anisotropy gradient. Specifically, for the two spins ($S_L$ and $S_R$) neighboring the spin $S_0$ exactly on the wall central plane, the damping torque on $S_L$, resulting from the anisotropy term, is smaller than that on $S_R$, leading to the fact that $S_L$ deviates more from the easy axis than $S_R$ does. As a result, the damping torque ($\mathbf{\Gamma}_R^d$) on $S_0$ from the exchange interaction with $S_R$ is larger than the damping torque ($\mathbf{\Gamma}_L^d$) from $S_L$, resulting in the net damping torque $\sim (\mathbf{\Gamma}_R^d + \mathbf{\Gamma}_L^d)$ which efficiently drives the wall motion toward the low anisotropy region. Undoubtedly, the net driving torque increases with increasing $\Delta K$, which significantly enhances the wall motion speed, as clearly shown in Fig. 2(d) where the wall position $z$ as a function of time $t$ at several $\Delta K$ is plotted.

In order to uncover the underlying physics more clearly, we calculate the wall motion speed from the perspective of energy conservation. Considering the wall motion as a consequence of wall energy dissipation, one notes that the wall energy $E_{DW} = 2(2JK_c)^{1/2}$ where $K_c$ is the anisotropy constant on spin $S_0$,[31] noting $K_c > K_0$ if $\Delta K > 0$. The Rayleigh dissipation function $R = \alpha\rho \cdot \int \dot{\mathbf{n}}^2 dz/2$ is introduced to describe the energy dissipation of the whole lattice,[32] where $\rho = \mu_s/\gamma a$ is the density of staggered spin angular momentum per unit cell and $\dot{\mathbf{n}}$ represents the derivative with respect to time. Subsequently, according to the classical mechanics, we obtain:

$$dE_{DW}/dt = -2R, \qquad (3)$$

A necessary mathematical substitution leads to the wall velocity $v$ (see Appendix for details):

$$v = -\frac{\gamma \lambda^2 \Delta K}{\alpha \mu_s}, \quad \lambda = a\left(\frac{J}{2K_c}\right)^{1/2}, \tag{4}$$

where $\lambda$ is the domain wall width.

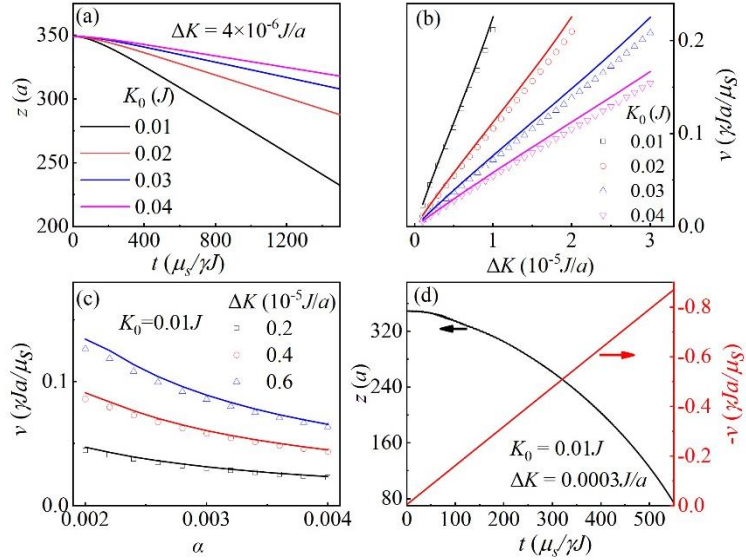

Fig.3. (color online) (a) The position of the domain wall as a function of time for various $K_0$ for $\Delta K = 4 \times 10^{-6}$ $J/a$. The simulated (empty points) and calculated (solid lines) velocities as functions of (b) $\Delta K$ for various $K_0$ and $\alpha = 0.002$, and (c) $\alpha$ for various $\Delta K$ for $K_0 = 0.01$ $J$. (d) The simulated domain wall position and velocity as functions of time for large $\Delta K = 0.0003$ $J/a$.

Eq. (4) suggests that the wall velocity depends not only on anisotropy gradient $\Delta K$ but also anisotropy constant $K_c$. It implies that a large anisotropy $\Delta K$ makes the wall motion fast but a larger $K_c$ does the opposite. This result is confirmed from our simulations. Fig. 3(a) presents the simulated wall position $z$ as a function of $t$ for various $K_0$ at $\Delta K = 4 \times 10^{-6}$ $J/a$, noting that a large $K_0$ also implies a large $K_c$. Certainly, an enhanced anisotropy is expected to suppress the domain wall motion seriously, as shown in Fig. 3(a) and Fig. 3(b). In Fig. 3(b), the simulated speeds (empty dots) and calculated ones (solid lines) from Eq. (4) are plotted together for a comparison, demonstrating the good consistence between the simulation and model on one hand and on other hand revealing the fact that the anisotropy magnitude ($K_0$ or $K_c$) and anisotropy gradient ($\Delta K$) play the opposite roles in controlling the domain wall motion, enabling an accelerated and decelerated wall motion respectively.

The effect of the damping constant $\alpha$ on the wall velocity is also investigated, and the corresponding results are shown in Fig. 3(c) which gives the simulated and calculated velocities as functions of $\alpha$ for various $\Delta K$ for $K_0 = 0.01J$. It is noted that an enhanced damping term always lowers the wall mobility, and the velocity decreases with the increasing $\alpha$. Furthermore, the wall width is simply considered to be a constant during the wall motion, which well describes the case of small $\Delta K$. However, for large $\Delta K$, the wall width is expected to increases with the wall motion toward the weak anisotropy side, resulting in the additional acceleration of the wall motion. In order to better understand the acceleration behavior, we investigated the wall motion under huge anisotropy gradient $\Delta K = 0.0003J/a$. The simulated wall position and velocity are presented in Fig. 3(d), which clearly shows that the velocity linearly increases with $t$.

Furthermore, the 90° AFM domain wall could be available in real materials such as CuMnAs using the current polarity and may play an essential role in future applications.[33] Here, the dynamics of the 90° AFM wall is also investigated by introducing the cubic anisotropy $K(S_x^2 S_y^2 + S_y^2 S_z^2 + S_z^2 S_x^2)$ and additional hard axis anisotropy $K_y S_y^2$ in the model.[34] In this case, the wall energy goes $E_{DW} = (JK_c/2)^{1/2}$, exactly a quarter of the energy of the 180° wall under the same parameters. However, the Rayleigh dissipation function $R$ of the 90° wall is also one-fourth that of the 180° wall, as explained in the appendix, resulting in the same velocity. As a matter of fact, this behavior has also been confirmed by the LLG simulations, and the corresponding results are not shown here for brevity.

*B. Lattice dimension and DM interaction*

So far, the anisotropy gradient driven AFM wall motion has been studied based on the one-dimensional model. However, the results and conclusions also apply to magnetic films and bulks, as shown in Fig. 4(a) which presents the simulated wall velocity as a function of $\Delta K$ for various lattice dimensions for $K_0 = 0.01J$ and $\alpha = 0.002$ with the in-plane exchange interaction $J$. All the curves well coincide with each other especially for small $\Delta K$, indicating the independence of the wall velocity on the lattice dimension. This phenomenon can be qualitatively understood from the energy landscape. Under a small $\Delta K$, the wall structure is almost the same as the static wall. In this case, the neighboring spins in every cross-section perpendicular to the $z$ axis still align antiparallel to each other, and the exchange energy is satisfied. Thus, the energy difference between neighboring unit cross sections only arises from the single-ion anisotropy term, which hardly be affected by the lattice dimensions, resulting in the invariant force density exerting on the wall.

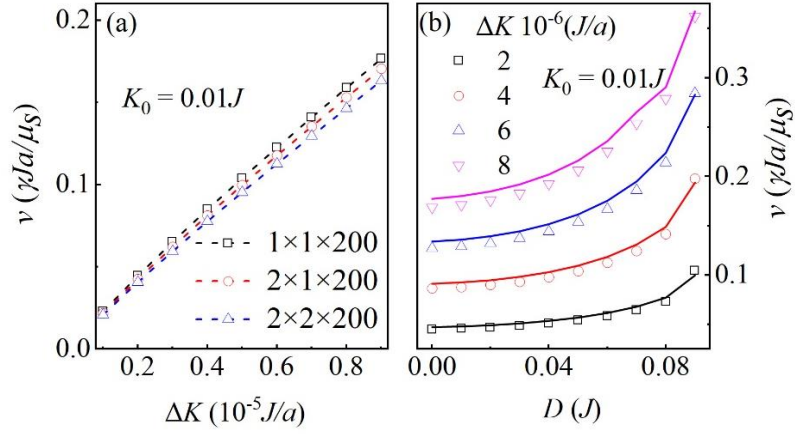

Fig.4. (color online) (a) The simulated domain wall velocity as a function of $\Delta K$ for various lattice sizes, and (b) The simulated (empty points) and calculated (solid lines) velocities as functions of $D$ for various $\Delta K$.

Moreover, we check the dependence of the domain wall dynamics on DM interaction which may be available in some realistic materials.[35,36] Here, the DM interaction energy $\Sigma_i \mathbf{D} \cdot (\mathbf{S}_i \times \mathbf{S}_{i+1})$ with $\mathbf{D} = D(0, 0, 1)$ is introduced in the one-dimensional model, and the LLG simulated results are presented in Fig. 4(b), where the $v(D)$ curves upon various $\Delta K$ are plotted. For a fixed $\Delta K$, $v$ gradually increases with the increase of $D$, indicating that the DM interaction enhances the wall motion, similar to the earlier report.[37,38] When the DM interaction is considered, the domain wall energy decreases to $E_{DW} = 2(2JK_c - D^2)^{1/2}$,[9,39] resulting in the increase of the wall width and the enhancement of the wall mobility. In detail, the wall width is increased to[40]

$$\lambda = \lambda_0 \left(1 - D^2/2JK_c\right)^{-\frac{1}{2}}, \tag{5}$$

with $\lambda_0 = a(J/2K_c)^{1/2}$.

In Fig. 4(b), the calculated velocity curves are also plotted with solid lines, which are well consistent with the simulated results.

## C. Antiferromagnet beats ferromagnet

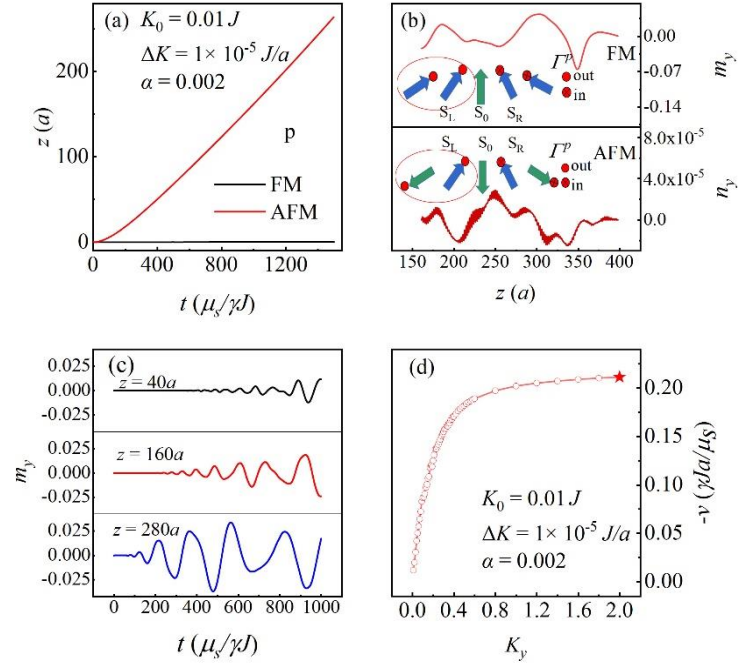

Fig.5. (color online) (a) The displacements of the domain walls of the AFM and FM systems as functions of time, and (b) the local $m_y$ (top half) of FM system and $n_y$ (bottom half) of AFM system, and the inserts are the schematic depictions of the precession torques, and (c) the evolution of $m_y$ at various positions, and the domain wall center is located at $z = 350a$, and (d) the simulated FM wall velocity as a function of $K_y$.

In this section, we discuss the physical mechanism for the ultrafast dynamics of an AFM wall under the anisotropy gradient through a detailed comparison with FM domain wall. Comparing with the AFM wall, the FM wall can hardly be driven by the anisotropy gradient, as clearly shown in Fig. 5(a) where presents the simulated wall displacements of the AFM system and FM system with the exchange interaction $-J$ for $\Delta K/K_0 = 1 \times 10^{-3}\ a^{-1}$ and $\alpha = 0.002$. For a FM wall, the precession torques on $\mathbf{S}_R$ and $\mathbf{S}_L$ from the gradient are antiparallel with and compete with each other, resulting in a strong spin fluctuation, as shown in the top half of Fig. 5(b) where the $y$ component of local spin $m_y$ is presented. Moreover, the spin fluctuation cannot be suppressed by the FM interaction and propagates along the nanowire from the domain wall center, as confirmed in Fig. 5(c) where gives the evolutions of $m_y$ at $z = 40a$, $160a$, and $280a$, noting that the domain wall center is located at $z = 350a$. As a result, the excitation and propagation of spin fluctuation act as dominating energy dissipations since the precession torque is much larger than the damping torque, resulting in the very low driving efficiency of the FM wall.

On the other hand, for the AFM system, the high mobility of the AFM wall is obtained, due to the fact that the spin fluctuation is significantly suppressed by the AFM interaction, as shown in the bottom half of Fig. 5(b) where presents the $y$ component of the staggered magnetization $n_y$ of the system and the depiction of the precession torque acting on the AFM spins. Thus, this work unveils another origin for ultrafast dynamics of an AFM system. However, the weak spin fluctuation also slightly speeds down the motion of the AFM domain wall, resulting in the noticeable discrepancy between the simulations and analytical calculations especially for large $\Delta K$ which significantly enhances the spin fluctuation, as shown in Fig. 3.

In addition, a high mobility of the FM wall is expected and also confirmed in our simulations when the spin fluctuation is suppressed by introducing additional hard-axis anisotropy $K_y S_y^2$ in the model. Fig. 5(d) shows the simulated FM wall velocity as a function of $K_y$. With the increase of $K_y$, the spin fluctuation is gradually suppressed, and the velocity increases to a saturation value approximates to that of the AFM wall.

## D. Brief discussion

While we are pursuing the faster, the smaller, the denser for spintronic devices, the domain wall motion efficiently driven by low energy-consuming can be always the core concern for spintronic study. Thus, finding new energy-saving methods of controlling the domain walls in antiferromagnet, regardless of it is metal or insulator, is still of great importance for future applications.

In this work, anisotropy gradient has been clearly revealed to efficiently drive the AFM domain wall motion. Experimentally, anisotropy gradient could be easily obtained through tuning electric voltage on particular heterostructures, and drive the AFM domain wall motion at a considerable speed. Specifically, for $\Delta K/K_c = 1 \times 10^{-4}\ a^{-1}$ and $\alpha = 0.002$, the speed of the wall for NiO with the exchange stiffness $A \approx 5 \times 10^{-13}$ J/m, $a \approx 4.2$ Å, $\mu_s \approx 1.7\ \mu_B$ ($\mu_B$ is the Bohr magneton) is estimated to be $\sim 100$ m/s. More importantly, the proposed modulating method is expected to be more energy-saving and with less Joule heat, comparing with those electric current driving methods. Thus, our work unveils a promising method of controlling AFM wall, and does provide useful information for future spintronic applications.

## IV. Conclusion

In summary, we have studied the AFM wall motion under the anisotropy gradient based on the one-dimensional model. The domain wall velocity depending on the gradient magnitude and intrinsic physical parameters are simulated based on the LLG equation and also calculated theoretically based on the energy dissipation theorem. The domain wall structure is rather robust for small gradient and moves at a constant velocity, while accelerates for large gradient due to the enlargement of the domain wall width during its motion. The domain wall mobility is independent of the lattice dimension and wall type (180° or 90°), while the mobility is enhanced by the DM interaction. Moreover, the physical mechanism of the AFM wall dynamics faster than FM wall dynamics is also qualitatively explained. This work unveils a promising strategy for controlling the AFM domain walls, benefiting to future spintronic applications.

**Acknowledgment**

We sincerely appreciate the insightful discussions with Zhengren Yan and Huaiyang Yuan. The work is supported by the National Key Projects for Basic Research of China (Grant No. 2015CB921202), and the Natural Science Foundation of China (No. 51971096), and the Science and Technology Planning Project of Guangzhou in China (Grant No. 201904010019), and the Natural Science Foundation of Guangdong Province (Grant No. 2016A030308019).

**Appendix: The derivation of the AFM wall velocity**

Following the earlier works, we use domain wall ansatz $\tan(\theta/2) = e^{(z-q)/\lambda}$ for 180° wall and $\tan(\theta) = e^{(z-q)/\lambda}$ for 90° wall with $q$ is the coordination of the wall center. The wall spins are restricted in the x-z plane ($\varphi = 0$), and the local staggered magnetization is given by:

$$\begin{aligned} n_x &= \sin(2\arctan(e^{(z-q)/\lambda})) \cdot \cos(\varphi) \\ n_y &= \sin(2\arctan(e^{(z-q)/\lambda})) \cdot \sin(\varphi) \\ n_z &= \cos(2\arctan(e^{(z-q)/\lambda})) \end{aligned} \tag{A1}$$

for 180° wall, and

$$\begin{aligned} n_x &= \sin(\arctan(e^{(z-q)/\lambda})) \cdot \cos(\varphi) \\ n_y &= \sin(\arctan(e^{(z-q)/\lambda})) \cdot \sin(\varphi) \\ n_z &= \cos(\arctan(e^{(z-q)/\lambda})) \end{aligned} \tag{A2}$$

for 90° domain wall.

According to the classical mechanics, one obtains $dE_{DW}/dt = -2R$. The domain wall energy is calculated by $E_{DW} = 2(2JK)^{1/2}$ for 180° wall and $E_{DW} = (JK/2)^{1/2}$ for 90° wall with $K = K_0 + vt\Delta K$. Subsequently, $dE_{DW}/dt$ reads

$$dE_{DW}/dt = \begin{cases} 2\lambda v\Delta K & \text{for } 180°\text{ DW} \\ \lambda v\Delta K/2 & \text{for } 90°\text{ DW} \end{cases}. \tag{A3}$$

Furthermore, using the mathematical transformation

$$\dot{\mathbf{n}} = \frac{d\mathbf{n}}{dt} = \frac{d\mathbf{n}}{dz}\frac{dz}{dt} = \frac{d\mathbf{n}}{dz}v, \tag{A4}$$

$R$ can be described by

$$R = \begin{cases} \alpha\rho v^2/\lambda & \text{for } 180°\text{ DW} \\ \alpha\rho v^2/4\lambda & \text{for } 90°\text{ DW} \end{cases}. \tag{A5}$$

At last, the domain wall velocity is obtained by

$$v = -\frac{\gamma\lambda^2\Delta K}{\alpha\mu_s} \tag{A6}$$

both for 180° wall and 90° wall.

**Reference:**

1. R. Cheng, J. Xiao, Q. Niu, and A. Brataas, Phys. Rev. Lett. **113**, 057601 (2014).
2. E. Gomonay and V. Loktev, Low. Temp. Phys+. **40**, 17 (2014).
3. V. Baltz, A. Manchon, M. Tsoi, T. Moriyama, T. Ono, and Y. Tserkovnyak, Rev. Mod. Phys. **90**, 015005 (2018).
4. O. Gomonay, T. Jungwirth, and J. Sinova, Phys. Status Solidi RRL **11**, 1770319 (2017).
5. Marrows and Christopher, Science **351**, 558 (2016).
6. Z. Y. Chen, Z. R. Yan, M. H. Qin, and J. M. Liu, Phys. Rev. B **99**, 214436 (2019).
7. Z. Y. Chen, M. H. Qin, and J. M. Liu, Phys. Rev. B **100**, 020402 (2019).
8. T. Kampfrath, A. Sell, G. Klatt, A. Pashkin, S. Mährlein, T. Dekorsy, M. Wolf, M. Fiebig, A. Leitenstorfer, and R. Huber, Nat. Photonics **5**, 31 (2011).
9. W. C. Yu, J. Lan, and J. Xiao, Phys. Rev. B **98**, 144422 (2018).
10. O. Gomonay, T. Jungwirth, and J. Sinova, Phys. Rev. Lett. **117**, 017202 (2016).
11. P. Wadley, B. Howells, J. Železný, C. Andrews, V. Hills, R. P. Campion, V. Novák, K. Olejník, F. Maccherozzi, and S. Dhesi, Science **351**, 587 (2016).
12. Y. L. Zhang, Z. Y. Chen, Z. R. Yan, D. Y. Chen, Z. Fan, and M. H. Qin, Appl. Phys. Lett. **113**, 112403 (2018).
13. S. K. Kim, O. Tchernyshyov, and Y. Tserkovnyak, Phys. Rev. B **92**, 020402 (2015).
14. S. Selzer, U. Atxitia, U. Ritzmann, D. Hinzke, and U. Nowak, Phys. Rev. Lett. **117**, 107201 (2016).
15. Z. R. Yan, Z. Y. Chen, M. H. Qin, X. B. Lu, X. S. Gao, and J. M. Liu, Phys. Rev. B **97**, 054308 (2018).
16. K. Olejník, T. Seifert, Z. Kašpar, V. Novák, P. Wadley, R. P. Campion, M. Baumgartner, P. Gambardella, P. Němec, and J. Wunderlich, Sci. Adv. **4**, eaar3566 (2018).
17. S. Y. Bodnar, L. Šmejkal, I. Turek, T. Jungwirth, O. Gomonay, J. Sinova, A. Sapozhnik, H.-J. Elmers, M. Kläui, and M. Jourdan, Nat. Commun. **9**, 348 (2018).
18. X. R. Wang, P. Yan, J. Lu, and C. He, Ann. Phys-new. York. **324**, 1815 (2009).
19. A. Mougin, M. Cormier, J. P. Adam, P. J. Metaxas, and J. Ferré, Europhys. Lett. **78**, 57007

(2007).

20. T. Nozaki, T. Yamamoto, S. Tamaru, H. Kubota, A. Fukushima, Y. Suzuki, and S. Yuasa, APL Mater. **6**, 026101 (2018).

21. S. Z. Peng, S. Li, W. Kang, J. Q. Zhou, N. Lei, Y. G. Zhang, H. X. Yang, X. Li, P. K. Amiri, and K. L. Wang, Appl. Phys. Lett. **111**, 152403 (2017).

22. X. Li, K. Fitzell, D. Wu, C. T. Karaba, A. Buditama, G. Yu, K. L. Wong, N. Altieri, C. Grezes, and N. Kioussis, Appl. Phys. Lett. **110**, 052401 (2017).

23. Z. Wen, H. Sukegawa, T. Seki, T. Kubota, K. Takanashi, and S. Mitani, Sci. Rep. **7**, 45026 (2017).

24. X. Wang, W. Gan, J. Martinez, F. Tan, M. Jalil, and W. Lew, Nanoscale **10**, 733 (2018).

25. H. Y. Xia, C. K. Song, C. D. Jin, J. S. Wang, J. B. Wang, and Q. F. Liu, J. Magn. Magn. Mater. **458**, 57 (2018).

26. L. C. Shen, J. Xia, G. P. Zhao, X. C. Zhang, M. Ezawa, O. A. Tretiakov, X. X. Liu, and Y. Zhou, Phys. Rev. B **98**, 134448 (2018).

27. F. D. M. Haldane, Phys. Rev. Lett. **50**, 1153 (1983).

28. D. Hinzke and U. Nowak, Phys. Rev. Lett. **107**, 027205 (2011).

29. P. Yan, X. S. Wang, and X. R. Wang, Phys. Rev. Lett. **107**, 177207 (2011).

30. Y. G. Semenov, X.-L. Li, and K. W. Kim, Phys. Rev. B **95**, 014434 (2017).

31. K. M. Pan, L. D. Xing, H. Y. Yuan, and W. W. Wang, Phys. Rev. B **97**, 184418 (2018).

32. S. K. Kim, Y. Tserkovnyak, and O. Tchernyshyov, Phys. Rev. B **90**, 104406 (2014).

33. P. Wadley, S. Reimers, M. J. Grzybowski, C. Andrews, M. Wang, J. S. Chauhan, B. L. Gallagher, R. P. Campion, K. W. Edmonds, and S. S. Dhesi, Nat. Nanotechnol. **13**, 362 (2018).

34. H. H. Yang, H. Y. Yuan, M. Yan, H. W. Zhang, and P. Yan, Phys. Rev. B **100**, 024407 (2019).

35. K.-S. Ryu, L. Thomas, S.-H. Yang, and S. Parkin, Nat. Nanotechnol. **8**, 527 (2013).

36. S. Emori, U. Bauer, S.-M. Ahn, E. Martinez, and G. S. Beach, Nat. Mater. **12**, 611 (2013).

37. S. G. Je, D. H. Kim, S. C. Yoo, B. C. Min, K. J. Lee, and S. B. Choe, Phys. Rev. B **88**, 214401 (2013).

38. D. H. Kim, D. Y. Kim, S. C. Yoo, B. C. Min, and S. B. Choe, Phys. Rev. B **99**, 134401

(2019).

39. A. Qaiumzadeh, L. A. Kristiansen, and A. Brataas, Phys. Rev. B **97**, 020402 (2018).

40. Y. Yamane, Phys. Rev. B **98**, 174434 (2018).

**FIGURE CAPTIONS**

Fig.1. (color online) Illustration of a domain wall in antiferromagnetic nanowire under an anisotropy gradient.

Fig.2. (color online) (a) Profile of the domain wall at different times, and (b) the domain wall position and energy as functions of time for $\Delta K = 6\times10^{-6}$ $J/a$ and $K_0 = 0.01$ $J$. (c) A schematic depiction of torques acting on domain wall spins under the anisotropy gradient, and (d) the position of the domain wall versus time for various $\Delta K$ for $K_0 = 0.01$ $J$.

Fig.3. (color online) (a) The position of the domain wall as a function of time for various $K_0$ for $\Delta K = 4\times10^{-6}$ $J/a$. The simulated (empty points) and calculated (solid lines) velocities as functions of (b) $\Delta K$ for various $K_0$ and $\alpha = 0.002$, and (c) $\alpha$ for various $\Delta K$ for $K_0 = 0.01$ $J$. (d) The simulated domain wall position and velocity as functions of time for large $\Delta K = 0.0003$ $J/a$.

Fig.4. (color online) (a) The simulated domain wall velocity as a function of $\Delta K$ for various lattice sizes, and (b) The simulated (empty points) and calculated (solid lines) velocities as functions of $D$ for various $\Delta K$.

Fig.5. (color online) (a) The displacements of the domain walls of the AFM and FM systems as functions of time, and (b) the local $m_y$ (top half) of FM system and $n_y$ (bottom half) of AFM system, and the inserts are the schematic depictions of the precession torques, and (c) the evolution of $m_y$ at various positions, and the domain wall center is located at $z = 350a$, and (d) the simulated FM wall velocity as a function of $K_y$.